# Does a Fine-Tuned Universe Tell Us Anything About God?[*]

*Adam D. Hincks*[1,2,3,†]

[1]*David A. Dunlap Department of Astronomy and Astrophysics, University of Toronto, 50 St George Street, Toronto ON M5S 3H4, Canada*
[2]*University of St. Michael's College, University of Toronto, 81 St. Mary Street, Toronto ON M5S 1J4, Canada*
[3]*Specola Vaticana (Vatican Observatory), V-00120 Vatican City State*
[†]*Email address: adam.hincks@utoronto.ca*

*10 January 2024*

## Abstract

The apparent fine-tuning of several fundamental parameters that determine the properties of our Universe and make it hospitable to life is sometimes used as an argument for God from design. I review the concept of cosmic fine-tuning and critically examine the claim that God is its most probable cause. While not definitively repudiating this claim, I argue that it is potentially in tension with the more apophatic approach to God found in the Abrahamic traditions. I then offer a metaphysical analysis of the contingency of fine-tuning that situates it within the classical analogy of being that points to the Divinity.

## Introduction

Fine-tuning refers to the fact that several very fundamental properties of the cosmos appear to be delicately balanced such that it is possible for life to exist. In a hypothetical universe where these properties are changed, in some cases only slightly, it is very difficult—indeed probably impossible—to envisage how living organisms of any type could exist. These striking coincidences that enabled the emergence of life are often used in arguments for why the Universe must have a divine Creator: in brief, the fine-tuning is construed as the result of a purposeful choice of an intelligence that is identified with God. Unsurprisingly, this sort of argument is not without its difficulties and there are naturalistic counter proposals to explain the fine-tuning. My goal in this paper is not, however, to dwell on the technical merits of probabilistic arguments for God based on fine-tuning but rather to look more deeply at the metaphysics of fine-tuning *per se* as a clue for how it might be related to the Abrahamic doctrine of creation. I shall propose that fine-tuning provides a compelling entry point to the classical *analogy of being* by which the created order gestures towards God.

To reach this point, I shall begin with a brief review of the fine-tuning phenomenon and how it has been leveraged to argue for a Creator, and then proceed to describe some of the difficulties with establishing the exact degree of fine-tuning together with its possible naturalistic explanations that have ramifications for a probability-based argument for God. I shall also explain how this sort of theistic argument, which leans on empirical observations, may be in tension with traditional philosophical approaches to God, as represented by Thomas Aquinas. The probabilistic approach runs the risk of being, in the language of Kathryn Tanner, "not radical enough" (1988, 46). To explore how

---

fine-tuning might better be related to theology, I first analyze how fine-tuned cosmic properties fit into empirical science itself, using the distinction between "classical" and "statistical" investigations proposed by Bernard Lonergan. This allows me, finally, to relate fine-tuning to Erich Przywara's notions of creaturely actuality and arrive at the analogy of being.

**The Phenomenon of Cosmic Fine-Tuning**

Today, we have a remarkably good understanding of how the Universe has expanded and evolved over its fourteen billion year history. One of the most important empirical tools we have for this is the *cosmic microwave background* (CMB), the glow of the primordial material of the Universe that was emitted about 400,000 years after the big bang. This is very early on in cosmic history, when the Universe was only 0.003% of its present age. There were no stars or galaxies, just a hot gas of roughly 3,000° C, which was glowing as any hot material does. We literally see that glow, but in microwaves rather than visible light because its wavelength has been stretched by a factor of around 1,000 due to the expansion of the Universe. The glowing gas was the material that collapsed into stars hundreds of millions of years later, but in this primordial epoch it was almost perfectly smooth everywhere, with variations in its density typically being only one part in 100,000. (To get a sense of scale, that is roughly how much a loud machine causes the density of air to oscillate. You need a very sensitive instrument, like your eardrum, to detect these tiny changes in pressure.) Still, these variations in density were extremely important, because where the Universe was slightly more dense, there was slightly more gravity, allowing that region to attract the surrounding material so that eventually stars and galaxies would form hundreds of millions of years later. We often call them the *seeds of structure* because they were the initial conditions for the rich distribution of stars and galaxies today.

   Now, there are a few aspects of this situation that appear fine-tuned. First, the average density of the Universe—the average density around which you have these part-in-100,000 variations—is quite special. If it had been larger, the force of gravity would have caused the whole Universe to stop expanding very early on and it would have collapsed in on itself. For instance, if it had been just 1% denser one second after the big bang, the Universe would have collapsed, effectively coming to an end, less than seven minutes later. On the other hand, if it had been a bit smaller, the reduced gravitational attraction would have allowed the Universe to expand much more quickly, making it far too diffuse for stars, or any sort of complex structure, to ever form. To have a Universe that lasts as long as ours, the initial density had to be extremely fine-tuned.

   Second, it is surprising that the density of the Universe was the same everywhere to a part in 100,000. That may not seem strange, because intuitively one expects that any gas in a volume to be spread out evenly. However, it takes time for a gas to even out. When we observe the CMB, we are looking at a volume of the Universe that was roughly 100 million light years across, but the Universe was only 400,000 years old, and the maximum distance light could have traveled in this time was only about one million light years, or a hundred times smaller than the span of the total volume. (The reason light would have traveled a million rather than 400,000 light years is that space was stretching the whole time.) Now, nothing can move through space faster than the speed of light. That means that there was not enough time for the density at one end of the volume to smooth out across to the other end of the volume—it was more than 100 times too far away for it to have had causal contact with that region. On this reckoning, the CMB that we observe contains roughly 10,000 independent regions that were never in causal contact with each other. That implies that at the beginning of the Universe, somehow these 10,000 different locations started out with almost exactly the same density. There is no *a priori* reason that this coincidence should obtain.

Third, the amplitude of the tiny, part-in-100,000 variations in this average density also appear fine-tuned. If they were too small, they would not have sufficient gravity to pull enough matter together to make stars or similar structures: the expansion of the Universe would win out against their gravity and we would just have a Universe filled with uninteresting, diffuse gas. If they were larger, then the Universe would have produced too many stars. The stars would have been so crowded that any planet would have gotten knocked out of orbit around its star by other nearby stars, making its surface conditions too unstable for life to evolve (Adams 2019, §6.2). And if the density variations were even larger, the gas would have collapsed into black holes near the beginning of the Universe, and we would not have any stars or planets. Thus, the size of density variations in our Universe was in a sweet spot: not too big or too small.

In addition to these global properties of the Universe, there are many other examples of fine-tuning in how the Universe operates. For instance, stars require a balance between the force of gravity, which pushes everything towards the core of the star, and the outward pressure produced by the energy generated by nuclear fusion in the core. If nuclear reactions were too weak, then there would not be enough energy for the star to even turn on and produce the outward pressure, but if they were too strong, then the pressure would be so great that the star will just explode. It turns out that in our Universe, the strength of gravity, the strength of the electric force and the masses of electrons and protons all happen to have values that make this balance inside stars possible. Furthermore, it is not only the existence of stars that hinges on these particular parameters. Fig. 1, reproduced from Barnes (2012), shows some other key properties of our Universe that depend sensitively on these parameters, with the strength of the electric force on the *x*-axis and the ratio of the electron and proton masses on the *y*-axis. The diagram invites us to imagine how nature would behave if the electric force and the electron and proton masses had different values than they actually do. The actual values are shown with a cross near the bottom-left of the graph, and the different colored regions show how radically different nature would be if the values were different. Region 7 (yellow) has to do with the stability of stars that was just mentioned: in this region, stars cannot exist because the balance between the nuclear reactions and gravity does not obtain. Region 4 (pink) is where complex molecules cannot form because atoms do not stick together strongly enough. In this case you could never form the sorts of substances required for biology. In Region 1 (purple), not even the simplest element, hydrogen, can exist, so there would be no water (among other things). Each of the other colored regions shows a different catastrophic outcome. Only the small white region in the lower left of the diagram produces a Universe with stars, water and the elements of the periodic table required to make complex structures for life. The strength of the electric force needs to be quite small, and the electron needs to be more than 100 times lighter than the proton.

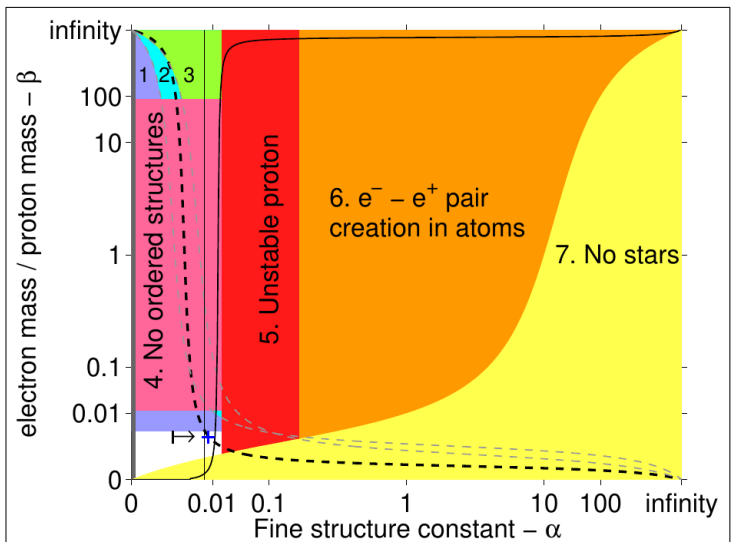

*Figure 1: An example of fine-tuning for two physical parameters: the ratio of the electron to proton mass (y-axis) and the strength of the electric force, α (x-axis). The true values of these parameters, as measured in our Universe, are indicated by the cross (with an arrow pointing to it) near the bottom left of the graph. The white region in which the cross lies shows which combination of parameters would allow the Universe to behave in such a way that life could still exist. The colored regions are where the parameters cause at least one difference in how the cosmos behaves that would be catastrophic for life. Graphic taken from Fig. 6 (left panel) of Luke A. Barnes, "The Fine-Tuning of the Universe for Intelligent Life," Publications of the Astronomical Society of Australia 29 (4): 529–64, 2013 © Cambridge University Press, reproduced with permission.*

Fig. 1 shows only two parameters that one can imagine changing, one on each axis. Altogether there are about a dozen such parameters (Adams 2019, Table 2); moreover, there are lots of different physical properties and processes that depend sensitively on their values. If it were possible to make a graph that did not have only two axes in two dimensions, like Fig. 1, but a dozen axes in a dozen dimensions to represent each parameter that is possible to vary, then you would still find only an extremely tiny white region where the parameters allow for a Universe capable of having stars and complex molecules and stable conditions. Furthermore, there are a number of other properties of our Universe that appear "selected" for life. To give just one example, we live in three spatial dimensions. There is no clear *a priori* why there could not be a different number of dimensions, but three dimensions appears to be the only configuration conducive to life. For instance, if there were more spatial dimensions, the orbits of planets around stars would not be stable.

A comprehensive exposition of all the known instances of fine-tuning in our cosmos is far beyond the scope of this paper, and I have only scratched the surface. Detailed scientific reviews of the subject can be found in Barnes (2012) and Adams (2019), and an excellent popular overview is provided by Lewis & Barnes (2016). However, the examples I have sketched above should be sufficient to convey the basic point, so I shall continue on to discuss the metaphysical implications.

**The Probabilistic Fine-Tuning Argument for God and Its Technical Difficulties**

Fine-tuning makes the Universe appear very special. It has just the right conditions after the big bang to grow into a cosmos suitable for complex structures to form, and it has constants of nature and other fundamental physical properties that seem to fall right in the sweet spot that make life possible. It should be stressed that many of the parameters involved in fine-tuning, such as the masses of particles or the strengths of forces, are not fixed by any known theory. There is nothing in our knowledge of physics that says that the proton must be 1836.15 times more massive than the electron. This ratio is a value we know only empirically, and the fine-tuning question arises because it is unclear why it and other fine-tuned parameters obtain in nature. But the fact that these particular values exist has been proposed as evidence for God, since it seems far more probable that fine-tuning is the result of a rational, purposeful choice of the Divinity rather than a fluke (see Simon 2021 for a review).

Probably the most familiar articulation of this sort of argument came from William Paley in the early nineteenth century. His analogy to someone coming across a watch lying in a field is well known: "[W]hen we come to inspect the watch, we perceive … that its several parts are framed and put together for a purpose, e.g. that they are so formed and adjusted as to produce motion … that, if the different parts had been differently shaped from what they are, of a different size from what they are, or placed after any other manner, or in any other order, than that in which they are placed, either no motion at all would have been carried on in the machine, or none which would have answered the use that is now served by it" (Paley 2006, 7). He considers this "invincible" evidence that the watch was deliberately created by a watchmaker (Paley 2006, 15), and, by analogy, the many instances of apparent design in the natural world are proof of an intelligent Creator. It is not difficult to understand why the fine-tuning discovered by modern physics and astronomy is seen by some as compelling evidence for the watch-maker argument. However, there is historical precedent for worrying about this line of reasoning. Paley's work leaned heavily biological examples of the intricacy of living creatures, but then Darwin appeared on the scene and provided another explanation. We should ask whether something similar could happen with cosmic fine-tuning, and in fact there is reason to take a closer look at just how improbable it really is. Let me raise three reasons that the Paley-style design argument must confront when it is applied to fine-tuning.

*The Difficulty in Quantifying Fine-Tuning*

Exactly how improbable it is that fine-tuning is the result of mere chance is challenging to determine for a couple of reasons. First, the size of the life-friendly region of parameter space is hard to pin down. Above, I emphasized its tininess, as represented by the white region in Fig. 1, but this size is based on current understandings of how nature works, and perhaps nature is more flexible than a diagram like this presumes. For instance, even if stars in our Universe seem to depend on a certain combination of parameters, investigations into the details of their nuclear reactions show that there may be alternative ways for stars to remain stable that expand the allowed parameter space (Adams 2019, §7), and it has been argued more generally that the parameters required for a life-friendly Universe are not as stringently constrained as is sometimes presented (Stenger 2011; see also Adams 2019, §10.6). Of course, this is contested (Barnes 2012). From my perspective, the actual degree of fine-tuning is still up for debate and further exploration, but denying that there is any fine-tuning at all (*à la* Stenger) is premature.

The second issue has to do with the range of *allowed* values. The axes in Fig. 1, which run from zero to infinity, present the most pessimistic and technically problematic scenario, for no matter how large your region of life-friendly values is, it will always be overwhelmingly small if you are drawing from an infinite number of possibilities. In fact, you cannot properly define probabilities at all, at least in a mathematical sense, if your parameter space is infinitely large. Now, it is possible that certain values are *a priori* disallowed, because, for instance, they would make physics internally inconsistent. For instance, the masses of particles are typically assumed to be smaller than the Planck scale at which gravity and quantum mechanics inevitably converge (e.g., Barr and Khan 2007; see also Barnes 2019, 1234–5). While this restriction of parameter space still allows for a huge range of values, it is at least finite. However, it is unclear whether the Planck scale is the correct cut-off point, exactly what span of values is actually possible for other parameters, and whether all allowed values have the same *a priori* probability (Adams 2019, 97–100; Hossenfelder 2021). Moreover, the foregoing presumes that the parameters are operating within the laws of physics that we have discovered in our Universe; if we allow for all possible laws of physics then determining the allowed range becomes yet thornier (Barr 2003, 146–8, 153–4).

Thus, given our current knowledge of physics we cannot definitively calculate how improbable fine-tuning actually is, and the lack of precision can be used to challenge the design argument (e.g., Hossenfelder 2021). This is not necessarily fatal to the theistic argument, as I shall briefly describe further on, but at the end of the day specifying the stringency of fine-tuning remains problematic at some level. We cannot calculate it in the same way that we can calculate how likely it is to draw a full house in poker.

*Fine-Tuning or Natural Process?*

Furthermore, the analogy of drawing random cards from a deck might not be a good one. Consider the Universe's initial density. As described above, it was just the right size, and it was incredibly uniform while also having minuscule variations of the right amplitude. It turns out that there is an elegant way that these seemingly special properties can be explained. If the Universe did not begin in such a smooth state, but with random, non-uniform densities everywhere, and then expanded extremely rapidly very early on, this would have smoothed everything out to the conditions that we observe in the CMB. This scenario is called *inflation*. It posits that the whole observable Universe came from a patch of space that was small enough that the density was already evened out inside of it right after the big bang. That tiny patch then expanded enormously—or *inflated*—during a fraction of the first second of the Universe, after which it continued expanding at a regular rate. The required amount of growth would be roughly equivalent to a proton becoming the size of our Solar System, all within about the first $10^{-34}$ seconds after the big bang. It may sound fantastical, but the physics that would have occurred during inflation is well understood. And crucially, the theory of inflation predicts an observable Universe with uniform density of the right size. It also explains the origin of the small density variations that are the seeds of structure. They are produced by quantum fluctuations during inflation, again quite well understood theoretically and whose properties match what we observe. Thus, inflation is widely considered to be the best theory of the early Universe that we have since it seems to explain our initial conditions so elegantly.

While the inflation paradigm is not without its issues, one of which I shall describe in the next section, it illustrates the possibility that parameters that *appear* to be fine-tuned might be the result of simple, physical mechanisms. If inflation may explain some of the initial conditions of the cosmological parameters, perhaps there are similar physical theories that will explain why other fine-tuned parameters take on the values that they do. The probabilistic design argument would then become impotent.

*The Multiverse*

Although inflation elegantly explains the initial conditions of our Universe, it does so at a cost, known as *eternal inflation*. For inflation to do the job of delivering our initial conditions, the period of rapid expansion is only required to last for a minuscule fraction of the first second after the big bang, but it turns out that theories of inflation typically have the curious property that the Universe goes on inflating indefinitely. They result in most of the Universe stretching as fast as I described earlier, where spaces as small as protons become as big as Solar Systems in a sliver of the blink of an eye. It is only in small *bubbles* or *pockets* that inflation stops and inside of which the subsequent expansion occurs at the "slower" pace that we observe in our Universe.[1] Exactly when and where these pockets appear is a random process, and it is precisely in these tiny pockets that a Universe like ours can be born: tiny, that is, compared to the immense sea of inflating space, but still so large that our whole, observable Universe can be contained inside one of these pockets. In other words, the pocket is large enough that we cannot see beyond it—light has not had time to reach us from beyond the pocket—but if we could, we would notice that our little region of the cosmos was unbelievably minuscule compared to the rest of space which is expanding at breakneck pace. Moreover, we would see scattered here and there other little islands like ours that might also have life-friendly conditions.

The foregoing scenario is a version of the *multiverse* hypothesis: namely, that physical reality is enormously bigger than our observable Universe. However, since any observer in the multiverse can only see a limited part of this reality, practically speaking it is as though there were multiple universes separated from each other by vast stretches of space and time.

The multiverse may seem far-fetched, and indeed some cosmologists eschew it because it seems impossible to ever empirically confirm: it seems too speculative to be real science (Ijjas, Steinhardt, and Loeb 2014; 2017). On the other hand, if inflation is the correct explanation for the initial conditions of our observable Universe, then the multiverse does seem to be a natural consequence (Guth, Kaiser, and Nomura 2014; Guth et al. 2017). Another attraction of the multiverse hypothesis is that it might explain not only the smoothness at the beginning of our Universe but also the fine-tuning of the constants of nature. In some theoretical frameworks, when a pocket of the multiverse stopped inflating, the numerical values of physical constants inside it would be randomly generated. In most pockets, you would get values that are completely unsuitable for the formation of stars and planets and life: you would end up in one of the problematic colored regions in Fig. 1. But if the multiverse extended far enough, then inevitably some of the pockets would happen to have just the right combination of values. So, even if the odds of a life-friendly Universe are minute, you would have enough different island universes that in at least some of them, you would get lucky. We would live in one of the lucky pockets where the strength of the electric force, the masses of protons and electrons, and so on, were all suitable for life to evolve. As some physicists have emphasized, this is reminiscent of the theory of evolution (e.g., Weinberg 2007, 39–40; Hawking and Mlodinow 2010, 165). Before Darwin, people could not fathom how complex organisms could come to exist through any natural process, but then the combination of random mutations with the principle of natural selection turned out to elegantly explain it. The complexity of our Universe would analogously be the result of random selection.

---

[1] In my experience, the term "bubble" is more widespread, but I prefer "pockets". It also has the virtue that it is used by Alan Guth, one of the pioneers of inflationary theory. See, for instance, Guth (2007).

To return to the poker analogy, the odds of getting a royal flush are one in six-hundred fifty thousand. If you were only dealt a single hand, this is very unlikely. But in the multiverse framework, the hands are being dealt continuously, everywhere, and so before long someone, somewhere gets a royal flush. Unfortunately, the poker analogy breaks down again when it comes to the question of hard numbers. One would like to be able to calculate precisely how likely our Universe is within the multiverse, just as one can calculate the odds of a poker hand. We would like to know how special we really are compared to all the other pockets. However, it turns out that there is not a straightforward way to answer this question. Technically this is called the *measure problem*: we do not know the "measure" for determining the odds. Thus, while the multiverse is an attractive explanation of fine-tuning for many people, it suffers from the problem that we cannot precisely say how likely or unlikely our pocket actually is. As we saw above when discussing the difficulty in determining how "small" the region of life-friendly parameter space is, it seems that our pocket of the multiverse must be "special" in some real way, but nailing this down with precision is elusive.

*Assessment of the Technical Challenges of a Probabilistic Argument*

The technical challenges presented above regarding a fine-tuning argument for God may not be insuperable. Even if nobody really knows how to calculate the exact probabilities involved in fine-tuning, proponents of the theistic argument argue that one only needs to establish that odds of the parameters being random are so small that on the balance of probability, it is much more likely that the fine-tuning comes from God rather than brute randomness (Barnes 2019; c.f., Swinburne 2010, 54ff). It has been argued that this approach, which one can do within the framework of Bayesian probability theory, is arguably at home with the way that scientists actually go about science. Namely, most scientists have *a priori* expectations about nature that are not precisely quantified, and yet are widely considered reasonable and work in practice (Lewis and Barnes 2016, 282–88).

Furthermore, it seems far-fetched to expect that the precise values of all of the fine-tuned parameters will eventually be apodictically derivable from pure theory, rather than being values that can only be empirically measured and therefore undetermined *a priori*. That is, it seems like wishful thinking to posit that fine-tuning is an illusion that will evanesce once a "theory of everything" proves that the value of these parameters *must* be what they are. While it may be true that inflation can explain away some of the fine-tuned parameters—though perhaps at the "cost" of a multiverse—there is no evidence that all parameters can be explained similarly. Indeed, the very hallmark of modern science is its empiricism: unlike in the old, Aristotelian approach, you cannot explain how nature works exclusively through deductive reasoning, but you need to observe and measure how it actually behaves. It seems reasonable to assume that some elements of physics and cosmology, such as constants of nature, will always remain brute facts, as far as science is concerned, that must simply be measured (*pace* Stenger 2011, 234). This assumption lies at the heart of the modern scientific project (see Hincks 2018, 333). Thus, I do not think that opponents of fine-tuning arguments are on firm ground when they appeal to future theories that will somehow show that our Universe's parameters are the only ones that are possible.

Finally, even though the multiverse is sometimes played as a trump card against the theistic argument, it yields some troubling possibilities. For instance, it has been argued that the multiverse would be far more likely to produce an isolated human brain complete with memories and the illusion that it lived in a full universe, than it would be to produce a full-fledged, finely tuned universe almost 100 billion light years across. The existence of multiverses containing more of these so-called *Boltzmann brains*—named for the nineteenth century physicist whose assistant inspired the idea (see Carroll 2021, 8–10 and references therein)—than proper universes is so bizarre that some think it deals a serious blow to the multiverse hypothesis (Lewis and Barnes 2016, 322).

In sum, it seems to me that the probabilistic argument for God from fine-tuning is a mixed bag. On the one hand, naturalistic responses can be leveled against it: the ambiguities in pinning down the probabilities, the hope of physical explanations for fine-tuning, and the possibility a multiverse will provide a simpler explanation. On the other hand, theistic responses to the naturalistic proposals are available that are extremely compelling to some people. After all, almost everyone agrees that fine-tuning is *striking* in some sense, and it seems like a short step to make the obvious link to the traditional, Abrahamic notion of God as the source of the purposefulness of the created order. Is it thus still possible to mount a convincing, probabilistic argument for God from design based on fine-tuning, despite the technical challenges I have outlined in this section? While I think there may be merit in pursuing this sort of argument, there is an important issue with this overall approach that I now wish to address.

**Probabilistic Fine-Tuning Arguments and the Theological Tradition**

The potential problem with probabilistic theistic arguments is their dependence on the results of empirical science. It places them in tension with the theological tradition that avoids treating God as the object of empirical investigation. The absolute transcendence of God is one of the cornerstones of Abrahamic theologies, which tend to insist that the apophatic approach to God never be far from view. In Western Christianity this was most famously articulated by the Fourth Lateran Council (1215): "For between Creator and creature no similitude can be expressed without implying a greater dissimilitude" (Denzinger 2010, #806). In other words, any framework that we use to talk about God must immediately be qualified as inadequate. Even though there may be true analogies for talking about God, the "similitude" that the analogy expresses will always be outweighed by its "dissimilitude". God's act of creation is no exception.

We can take Thomas Aquinas, who lived soon after Lateran IV, as our representative of this tradition. Thomas is careful not to confuse God's act of creation with the sort of "creation" that occurs in the regular course of nature (as when, for instance, a tree is "created" from an acorn or a star is "created" from a cloud of dust). He goes so far as to consider them equivocal: "[T]he universal production of being by God is neither motion nor mutation, but a certain simple emanation. And so 'to become' and 'to make' are used equivocally in reference to this universal production of things and in reference to other productions" (Thomas Aquinas 1963, Book VIII, Lect. 2, 974). By "motion" and "mutation" here, one should understand the sorts of physical processes that are studied by the empirical sciences. Thomas is emphatically denying that God's act of creation is of this sort.

The potential danger, therefore, with arguing that God is the "most probable" explanation of fine-tuning is that it treats of God at the same level as the empirical world. Even if someone making this argument frames God as being apart from the Universe in some way, God remains one option to be weighed against other options and there is the risk that his transcendence is lost. Indeed, as Kathryn Tanner has persuasively argued, if God is the Creator of everything as proposed by the Abrahamic traditions—if he is "the immediate source of being of every sort"—then his transcendence must be thoroughgoing and absolute (Tanner 1988, 46). This sits in possible tension with the notion that he is "more likely" than another cause, for even if he is "overwhelmingly" more likely, he is still being conceived of in contrast to other possible causes. As Tanner puts it: "Divinity characterized in terms of a direct contrast with certain sorts of being or with the world of non-divine being as a whole is brought down to the level of the world and the beings within it in virtue of that very opposition: God becomes one being among others within a single order" (Tanner 1988, 45). Thus, probabilistic design arguments may not go deep enough to fully arrive at the Abrahamic notion of God—and probably also with the conceptions of the Divine that are found in other major religious traditions (see Hart 2013, 27ff).

A consideration of the phenomenon of miracles can elucidate this point. It is beyond the scope of this paper to rigorously define or analyze miracles, but a good working definition is that miracles are phenomena that exceed or subvert nature's regular course so as to demand a supernatural explanation (e.g., Thomas Aquinas 1912–1936, I.110.4). Probabilistic fine-tuning arguments operate in a similar spirit, since the claim is that nature *per se* is not an adequate explanation for the conjunction of parameters necessary for a life-friendly universe. However, it is noteworthy that traditional demonstrations for God's existence and about his attributes tend to be based on metaphysical arguments rather than an appeal to miraculous phenomena. In the First Part of Thomas's *Summa Theologiae* which deals with God and the created order in general, including his famous five proofs for God's existence in Question 2 and his treatise on creation in Questions 44–47, miracles are only treated of in any detail in a few articles towards the end, comprising a minuscule fraction of the total text.[2] For Thomas, miracles play an epistemic role that is posterior to the question of God. God is knowable in a general way through "natural reason", whereas miracles are a "gratuitous grace" that aid the "supernatural knowledge"—i.e., faith—that is relevant for salvation (1912–1936, II-II.178.1). He tends, therefore, to speak of miracles as "for confirming faith [*ad fidei confirmationem*]" (e.g., 1888, II–II.178.1, *ad* 5; 1965, VI.2, *arg* 9) and "for showing [*ad ostendendum*]" or "for manifesting [*ad manifestandum/manifestationem*]" God's power and presence (e.g., 1888, III.43.1,3,4; 1961, III.9.10;

---

[2]Miracles are not treated per se until Question 105, Articles 7 and 8. Before this they are only mentioned in objections and responses to objections with only one passing exception (I.104.4). Subsequent to Q. 105, they are discussed chiefly with reference to angelic powers (I.110.4, I.114.4). The foregoing is based on a search for "miracl*" in the *Index Thomisticus* (Busa and associates, n.d.).

1965, VI.1, *ad* 4) rather than providing demonstrations of his existence in the strict sense.[3] This may help explain why probabilistic fine-tuning arguments do not have universal appeal. Those who accept that there is a God may find that fine-tuning is a powerful confirmation and an aid to faith, but it is not always sufficient to convince others. As Thomas himself points out, it is possible for us to mistakenly call something a miracle if we do not fully understand how nature works (1912–1936, I.110.4), and this is precisely the sort of counter-argument that can be made against probabilistic fine-tuning arguments, as we saw above.

Furthermore, as explained above, God's act of creation is not a physical process and is thus not empirically accessible. It therefore falls outside of the category of the miraculous, since by definition miracles are always understood with reference to nature, and the act of creation is the ground of nature rather than an operation within nature (Thomas Aquinas 1912–1936, I.105.7, *ad* 1). To put it another way, the God–world relation, which is what the Abrahamic doctrine of creation articulates, is metaphysically prior to anything that natural sciences can investigate. Of course, this does not mean that there might be miracle-like clues, such as fine-tuning, within nature that point to a reality that transcends nature. I am not claiming that probabilistic analyses of fine-tuning are a theological dead-end—after all, miracles can be compelling signs in aid of religious faith, as Thomas certainly thinks and the broader Catholic tradition affirms[4]—but rather that they would not provide direct access to the work of creation *per se*.

Moreover, probabilistic design arguments are not the only option. Thomas is sometimes credited with including an argument from design among his "five ways" for demonstrating God's existence (e.g., Ratzsch and Koperski 2022, §2), but it is quite different from Paley's:

> The fifth way is obtained from the governance of things. For we see that certain things that lack cognizance—namely, physical bodies—act for an end. This is evident from the fact that they always or very frequently act in the same way, such that they attain that which is best. It clearly follows that they arrive at this end not by chance but by an intention. For things that do not have cognizance do not tend to an end unless directed by something that is cognizant and intelligent, like an arrow from an archer. Thus there is something intelligent by which all physical things are ordered to an end, and we call this God (Thomas Aquinas 1888, I.2.3).

Thomas's approach is almost the opposite of Paley's: rather than appealing to the *improbability* of the world's constitution it relies on the *regularity* of the laws of nature and the fact that physical processes operate in intelligible ways with determinate ends. Now although fine-tuning appears necessary for our universe to be the way it is, in isolation, fine-tuning does not establish that laws of nature are regular or intelligible. It may be the result of regular laws of nature, and it may be part of a larger process that produces a life-friendly universe, but in and of itself, fine-tuning is just a set of facts. It is not clear whether fine-tuning is even relevant to an argument like Thomas's, which is about the general intelligibility of natural processes, not a particular set of facts.[5]

---

[3]Here, and whenever Latin editions of Thomas Aquinas are cited, translations are mine.

[4]In response to certain fideistic ideas, the First Vatican Council (1870), insisted that miracles play an integral role in the dynamic of faith. The Council document *Dei Filius* explains that faith, as a divine gift, has an "interior" aspect, but it is integrated with "exterior proofs", since "miracles and prophecies … are most certain signs of the divine revelation, adapted to the intelligence of all men" (Denzinger 2010, #3009; c.f., ##3010, 3033–34). The context here is revelation rather than natural theology—the latter of which is still considered, though, a sure path to knowledge of God (Denzinger 2010, ##3004, 3026). It is natural theology that is pertinent to this essay.

Still, the laws of nature are discovered by empirically studying particular sets of facts, so we should ask whether there is a relationship between the fine-tuning of physical parameters (used for Paleyesque arguments) and the regularity of nature (used for the Thomistic approach). Here we can benefit from the analysis of the twentieth century Jesuit philosopher Bernard Lonergan.

**Fine-Tuning and Emergent Probability**

Lonergan noted there are two complementary ways to find intelligibility in nature (Lonergan 1992, 102ff and 126ff). The first is via laws that use mathematical formulas to describe how the world regularly behaves, like thermodynamics or Einstein's theory of gravity. He calls this a "classical investigation" of nature. The second is via the patterns of actual events and facts that repeatedly occur in nature. For instance, an astronomer can measure where a planet is every night and make a table of positions. From that table she can figure out what the planet's average motion is. Or again, a geneticist can look at the offspring of a particular species of animal and figure out how often certain traits appear. Lonergan calls this second type of empirical activity "statistical investigation". He then points out a profound "complementarity" between classical and statistical investigation. On the one hand, the laws of nature discovered by classical investigation are based upon the average occurrence of events or facts that are uncovered by statistical investigation. For instance, the laws of planetary orbits were derived by abstracting equations from the planets' average motions observed over many nights; or again, the laws of genetic inheritance are derived by abstracting from the rates at which physical traits occur to an explanation of how traits are transmitted. On the other hand, statistical investigation tests the validity of classical laws by verifying whether the measured events are configured and ordered in the manner predicted by the theory. Thus, the two go together: laws are abstracted from the behaviors of events, and the behaviors of events occur in patterns governed by the abstract laws. Close attention to the pattern of the behaviors also allows classical laws to be refined. For instance, in the early twentieth century, astronomers were puzzled when measurements of the orbit of Mercury deviated very slightly, but very reliably, from the predictions of Newton's theory of gravity.[6] When Einstein formulated his new theory of gravity, one of its first triumphs was that it cohered with the average behavior uncovered by the statistical investigation. The previous theory remained intact in the sense that Newton's equations provide a pretty good approximation to Einstein's equations in most regimes, but now the classical theory had been refined.

---

[5]The implicit premise in the argument of the Fifth Way is that purposeful behavior must ultimately be explained by an intelligent agent. Providing an argument for this premise would be beyond the scope of this paper. However, it is worth pointing out that today people frequently interpret Aristotelian final causality in an overly anthropomorphic manner, as though there were a quasi-conscious agent hidden inside anything operating for an end. Final causality, however, just means what Thomas says: that processes "are ordered to an end". There is no requirement for an immediate intentionality; a good modern example would be the least action principle in physics. Thomas's claim in the Fifth Way, though, is that this sort of principle requires an intelligent ground.

[6]Lonergan uses the example of the precession of Mercury's perihelion in a couple of places, though to illustrate different points. See Lonergan (1992, 728, 733).

Crucially, though, statistical investigation does more than confirm or refine classical laws: it carries scientific understanding beyond the merely classical to the more complex intelligibility of the world. In isolation, classical theories describe "systematic processes", in which a whole collection of data can be fully understood by the single application of an abstract law (Lonergan 1992, 71). But as any working scientist will tell you, data never perfectly match the predictions of abstract laws. Lonergan calls these deviations from classical predictions "statistical residues" (Lonergan 1992, 109ff, but esp. 117ff). The residues may arise from random measurement uncertainties—and we know from quantum theory that at a certain level of precision this randomness is truly irreducible. They can also occur when no single classical law provides a complete explanation of a phenomenon. For instance, the deviation of today's weather from yesterday's weather forecast may be explainable by some appropriate combination of thermodynamics and fluid dynamics and geophysics, but there is no systematic way to relate these classical frameworks that does not sensitively depend on intial conditions. Each day the relevant conditions are such that the combination and application of classical laws is different, and as a result weather prediction is never perfect. In "non-systematic processes" like weather, the ensemble of data cannot be explained by a single classical law, but requires an unpredictible conjoining of multiple classical explanations (Lonergan 1992, 72) and inevitably includes some irreducible randomness. Thus, non-systematic processes are intelligible, but their intelligibility is complex.

Moreover, "non-systematic process can be the womb of novelty" (Lonergan 1992, 75). It is possible for a pattern to emerge in statistical residues that are merely coincidental with regard to one theory or set of theories, but which can be explained by a completely new type of theory. The various branches of science exist because what is just a "coincidental manifold" of statistical residues to one science can be explained in a non-coincidental manner by a higher science (Lonergan 1992, 230). So, to use Lonergan's example (1992, 230, 281ff), all of the molecules in a biological cell behave according to chemical laws. However, there are certain very regular patterns of behavior within a cell that are not explained by chemistry. You could not use chemistry to predict *a priori* that a cell should have particular parts like a nucleus, a membrane or a Golgi apparatus. Nonetheless, all of these cellular parts are intelligible and require further explanations that chemical theory cannot provide: they require biological theory. When molecules are arranged in particularly precise and complex ways, they become not merely molecules but constituents of a living organism. Biology builds upon chemistry.

The latter example shows that in the interplay between the abstract regularity of the world and the deviations of actual events from that regularity, it is possible for new processes with new intelligibility to emerge. Lonergan calls this "emergent probability" (1992, 144ff, but esp. 280ff). Note that this concept is a metaphysical principle rather than a scientific theory. It is an explanation for why emergence is *possible* based on the complementarity between abstract intelligibility and concrete intelligibility, but it does not predict whether a higher system will *actually* emerge in any particular situation. Whether it does or not is an empirical matter that must be determined in the usual scientific way. To return to the example of biology emerging from chemistry, as far as I am aware, we are still ignorant about how that emergence happened. If there is a young planet with a chemical soup floating around, we are not sure whether life will inevitably form, or whether it is extremely unlikely, or whether it requires additional scientific ingredients about which we are currently unaware. Perhaps one day we shall know the answer. Lonergan's notion of emergent probability, though, simply explains the mutual open-endedness of classical and statistical intelligibility, of which the appearance of life is a particular instantiation.

Now, we can consider fine-tuning in the context of emergent probability. Fine-tuning is striking because a number of seemingly disparate aspects of the early Universe and of physics have conspired together to enable a universe like ours to develop. It has the quality of a statistical residue, since the physical conditions of the early Universe together with the values of the parameters of physics constitute a concrete pattern that is not predicted by the big bang theory *per se* nor by the structure of our (current) laws of physics *per se*. However, this pattern turns out to be just the right one such that particular, complex structures can emerge, like galaxies and stars and planets and a periodic table of elements that can form complex molecules and eventually life. Therefore, the science of physical cosmology, which provides a unified theory explaining the behavior of the Universe on its largest scales of space and time, may be the fruit of emergent probability. What to elementary physics appears merely as a "coincidental manifold" (Lonergan 1992, 230) of parameters may in fact be the basis for the intelligibility of our physical Universe as a whole. Now, as stressed above, although it appears that the emergence of the pattern called fine-tuning is unlikely in some real sense, it is difficult to quantify. Perhaps it will be discovered that the pattern was inevitable after all and is actually not an example of emergent probability. Perhaps it was a hugely improbably fluke. Perhaps there is a vast multiverse and we inhabit a lucky spot where the right pattern happened to occur. Like the question about how life first appeared on earth, this array of possibilities appears to me to be empirical in nature—and narrowing down the field is something that scientists are working hard at.

However, we might ask whether behind the empirical question about fine-tuning there is more going on. Is there any significance to the fact that fine-tuning is a pattern that is open to the emergence of further, complexly ordered patterns that indeed appear to be necessary for our very existence?

**God and the Contingency of the Actual**

To explore this question let me call upon another twentieth-century Jesuit philosopher, Erich Przywara, whose masterpiece, *Analogia Entis,* was only translated into English relatively recently (2014). Przywara analyses what it means to be a being in the world: how "creatureliness", as he puts it, is constituted. He characterizes it as an "immanent dynamic middle". Thus, not only is creaturely being "immanent", since it is actual and concrete, but its immanence is a "dynamic middle" that is "suspended" between two kinds of possibility (Przywara 2014, 216). On the one hand, what is actual or immanent is contingent on other states of affair. For instance, this essay that you are reading is immanent: it is actually before your eyes. But its being here is contingent on a whole host of factors: it is contingent on my having written it; it is contingent on me having thought about the topic of fine-tuning beforehand; it is contingent on my education; it is contingent on your computer screen or printer reproducing the text for you; and so on. On the other hand, what is actual or immanent is also the starting point for what will come next. The concrete is open to something new. This essay that you are actually reading is moving things forward: by its nature, it opens your understanding out towards a better understanding of fine-tuning—or perhaps equally it fails to do so. The point is that it is not a static actuality but one that lays out further possibilities. These possibilities are certainly conditioned: the ideas that the essay stirs up are dependent on its actual content and how well they are organized and communicated and how cogent they are. And so it will not move you towards just any old ideas. But exactly what ideas it will stir up in you are still to be seen. The ending point is not completely determined.[7]

Przwyara thus concludes that the actuality of creaturely being is "suspended" (as he puts it) between the possibility out of which it emerges and the end towards which it is opening up; relating this to traditional philosophical terms, he explains that "actuality (ἐνέργεια) [is] between dynamic possibility (δύναμις) and an inner end-directedness (ἐντελέχεια)" (2014, 216). This means that there is a paradox in the being of this world. Beings are actual, but their actuality emerges from possibility and moves towards possibility. They are suspended between possibilities, between non-actualities. In other words, their being—their "isness"[8]—is negatively qualified. Whenever there is a state of affairs, Przywara reasons, it "is not not" (2014, 207–8). It actually "is": but it only *is* contingently. It could *not* have been if the right conditions had not existed, and it is *not* what it still can be. It *is* by virtue of being *not not.* It is *not* not actual, because what was only possible has in fact been made actual. And it is not *not* actual, because what is still possible is not yet and its actuality obtains here and now (see Hincks 2020, 30). To link this back to Lonergan, "non-systematic process can be the womb of novelty" (Lonergan 1992, 75) because statistical residues reveal that classical laws are embedded in a world of "nots" that make room for newness to emerge.

The actual is therefore existentially striking. Aristotle famously observed that the world elicits wonder in us. We encounter a phenomenon, and we want to explain it: we want to know how it came to be and what it can become (Aristotle 2014, I.2, 982b12). In the framework of Przywara, we want to know why it *is not not*. In the framework of Lonergan, scientists are struck by phenomena and wonder if there is a pattern. They do a statistical investigation to discover the pattern. If they find one, they discern abstract principles in it. Now, the configuration of the early Universe and the concrete values of the constants of nature constitute a particularly striking pattern that stimulates wonder. We wonder where it came from and to what extent it was necessary for what came next.

When it comes to the God question, I propose that we miss something significant by jumping immediately from this wonder to the question of what came before or after fine-tuning. To do so is to head away from the *is* to study the *nots* between which the *is* is suspended. Aristotle may have said that wonder drives philosophical inquiry, but he goes on to claim that the goal of philosophy is to *remove* wonder. Once you understand the causes of something, you no longer wonder about it (Aristotle 2014, I.2, 983a15). But the Divinity, at least in the great religious traditions, is not something that removes wonder, as if you could arrive at a definitive understanding of his nature.

---

[7]Lonergan calls this "finality" (1992, 470ff): through emergent probability, new, concrete states of affair can arise that are not necessarily mechanistically predicted. Here Lonergan adds something to Przywara, or at least brings something forth that is only implicit in his writing. Przywara seems to have in mind simply the classical notion of form with a proper τέλος. Lonergan's notion of emergent probability, though, shows that not only can the actualized form develop according to its immanent τέλος, but that new τέλοι can emerge from a matrix of actual forms.

[8]My term, not Przywara's, as far as I am aware.

In this vein, Przywara's chief insight is that the suspension of the actual between possibilities does not resolve itself. The way we humans know is through eliminating possibilities: we only know that something *is* insofar as it *is not not*. We know by comparison, by contrast, by elimination. The actual *isness* of creatures in the Universe is never naked: creaturely being is always clothed with *nots*. But you cannot ultimately get the core notion of *is* from a *not*. The notion of isness, of being, must ultimately originate elsewhere (Przywara 2014, 216–37).[9] In more traditional philosophical language, the being we are familiar with—the being of this Universe—is contingent. That is what Przywara means by "not not". Thomas Aquinas's famous demonstrations for God's existence look at this contingency from various angles—from five angles, to be precise—and conclude that the world's contingency is only resolved in something that is actual *per se*, and "we call this God" (1888, I.2.3). The *is* in the *is not not* of creatures is only finally explained by the unqualified *is* of God. And since there is no *not* in the *is* of God, there is no termination of wonder, unlike in every other sort of investigation.

Thus, I propose that fine-tuning gestures towards God not primarily because he is an explanation for why it was possible for fine-tuning to arise and for the sorts of universe that fine-tuning produces, even if that is certainly true in the ultimate analysis of God's creative power. Rather, the more important clue is the actuality of the fine-tuning: the unqualified wonder that it *is*. I use the term "unqualified" quite deliberately, because practically everyone who learns about fine-tuning finds it striking and wonderful, but to look for explanations that head away from the actuality—that attempt to *qualify* it—is paradoxically to head away from God. It is to leave the *is* in search of the *not*. Now, that is in fact the correct endeavor for the natural scientist and perhaps for the philosopher as well. When your question is about God, though, it is the wonder, the strikingness, the isness, the being *per se* that is arguably most relevant because that is the closest analogy we have to what God is like: the analogy of being (see again Przywara 2014, 214–37).

**Conclusion**

My assessment of what fine-tuning says about God may appear quite modest. I have framed fine-tuning as a created actuality that has an intelligible pattern. It is actual, but where it comes from and where it heads is provisional, is suspended. The suspension is not resolved in itself and points to a transcendent resolution. There are plenty of other things in the world that point to God through this analogy of being—indeed, everything points to him. You do not need to go to the fine-tuning of the Universe to find it, for it is all around you. If you dwell on what is actual in any type of creature or pattern or situation, even in the most ordinary circumstances, you have there a portal to the Divine, even if it is "in a glass, darkly". To coin a phrase from Ignatian spirituality, you can find God in all things (see Ignatius of Loyola 1952, ##235–7). Now, I do not want to dismiss probabilistic arguments outright. But I reckon that they will be most successful insofar as they are framed in the deeper context of the thoroughgoing contingency of the world; otherwise they will be, to return to Tanner, "not radical enough to allow a direct creative involvement of God with the world *in its entirety*" (1988, 46 emphasis mine). For the Abrahamic claim is that God does not just orchestrate coincidences like fine-tuning but that the world depends on him in every respect; hence Lonergan can state that "the question of God is implicit in all our questioning" (2017, 101), since every instance of contingency (or "suspension") prompts questions in us about its origin that remain ultimately unresolved within the world of contingency.

---

[9]The foregoing is an extremely compact summary of what Przywara means by the analogy of being, but hopefully gets at its core. In the cited pages, see especially 219 and 235–7.

And yet, if fine-tuning is just one example of suspended actuality among many, it is certainly a remarkably striking example. Even if the analogy of being is in the world around us, and indeed within us, it can be hard to see just how remarkable created actuality is when we are busy living in that same world. You have to step outside of your ordinary routine and have a contemplative experience of the natural world, or notice how remarkable it was how a concrete decision in the past impacted your subsequent life, or reflect upon the peculiarities of your own culture that you are normally unaware of, to be struck by the contingency, indeed the strangeness, of the actual. When you have such moments as these in which you pause and see how created actuality is nested before and ahead in possibilities, you get a glimpse of that suspendedness. But immersed as we are in this matrix of "suspended", "dynamic middles" (Przywara 2014, 216), and therefore usually blind to it, there is an advantage of being able to look out at the cosmos, far from our own familiar world of existence, to see how created being is writ large. As far out in space and back in time that we can see, the actual, the *is* of creation, is starkly present.

**Acknowledgments**

The germ of this essay was a paper entitled "Cosmic Fine-Tuning and Divine Causality" that I presented at the symposium on "Understanding Our Place in the Universe: Beyond the Legacy of Stephen Hawking" in Jerusalem in 2019. I am grateful to the symposium organizers Eric Priest and Mary Ann Meyers (the latter of whom also provided useful feedback on this essay) and all the other participants for starting me thinking about the topic, and to the John Templeton Foundation for funding my participation in that symposium and my work on this essay. I developed my thoughts further for a 2023 lecture at the jointly sponsored by the Christianity and Culture Program at the University of St. Michael's College (at the University of Toronto) and the University of St. Michael's College Philosophy Group. I am grateful for funding received from the latter, and I thank Michael Vertin and Reid Locklin for the invitation and organization of the lecture. This essay is largely based upon that talk. Finally, I thank Stephen Barr, Vincent Strand, S.J. and an anonymous referee for helpful comments on a draft of this paper; Dr. Barr's suggestions in particular prompted me to add the paragraphs on miracles and guided their contents.

**References**

Adams, Fred C. 2019. "The Degree of Fine-Tuning in Our Universe — and Others." *Physics Reports* 807 (May):1–111.

Aristotle. 2014. *Metaphysics (Books 1–9)*. Translated by Hugh Tredennick. Loeb Classical Library 271. Cambridge, MA: Harvard University Press.

Barnes, Luke A. 2012. "The Fine-Tuning of the Universe for Intelligent Life." *Publications of the*

*Astronomical Society of Australia* 29 (4): 529–64.

———. 2019. "A Reasonable Little Question: A Formulation of the Fine-Tuning Argument." *Ergo, an Open Access Journal of Philosophy* 6 (20201214). https://doi.org/10.3998/ergo.12405314.0006.042.

Barr, Stephen M. 2003. *Modern Physics and Ancient Faith*. Notre Dame, Indiana: University of Notre Dame Press.

Barr, Stephen M., and Almas Khan. 2007. "Anthropic Tuning of the Weak Scale and of $m_u/m_d$ in Two-Higgs-Doublet Models." *Physical Review D* 76 (4): 045002. https://doi.org/10.1103/PhysRevD.76.045002.

Busa, Roberto, and associates. n.d. *Index Thomisticus*. Web edition (English version) by Eduardo Bernot and Enrique Alarcón. Accessed December 26, 2018. http://www.corpusthomisticum.org/it/index.age.

Carroll, Sean M. 2021. "Why Boltzmann Brains Are Bad." In *Current Controversies in Philosophy of Science*, edited by Shamik Dasgupta, Ravit Dotan, and Brad Weslake. Current Controversies in Philosophy. New York ; London: Routledge, Taylor & Francis Group.

Denzinger, Heinrich. 2010. *Compendium of Creeds, Definitions and Declarations on Matters of Faith and Morals*. Edited by Peter Hünermann, Robert Fastiggi, and Anne Englund Nash. 43rd ed. San Francisco: Ignatius Press.

Guth, Alan H. 2007. "Eternal Inflation and Its Implications." *Journal of Physics A: Mathematical and Theoretical* 40 (25): 6811–26.

Guth, Alan H., D. I. Kaiser, and Y. Nomura. 2014. "Inflationary Paradigm after Planck 2013." *Physics Letters B* 733 (June):112–19. https://doi.org/10.1016/j.physletb.2014.03.020.

Guth, Alan H., David I. Kaiser, Linde, Andrei D., Yasunori Nomura, and Charles L. Bennet. 2017. "A Cosmic Controversy." *Scientific Amercan: Observations* (blog). May 2017. https://blogs.scientificamerican.com/observations/a-cosmic-controversy/.

Hart, David Bentley. 2013. *The Experience of God: Being, Consciousness, Bliss*. New Haven: Yale University Press.

Hawking, Stephen, and Leonard Mlodinow. 2010. *The Grand Design*. New York: Bantam Books.

Hincks, Adam D. 2018. "What Does Physical Cosmology Say about Creation from Nothing?" In *Creation Ex Nihilo: Origins, Development, Contemporary Challenges*, by Gary A. Anderson and Markus Bockmuehl, 319–45. Notre Dame: University of Notre Dame Press.

———. 2020. "Natural Knowledge of Creation from Nothing." Toronto: Regis College and University of Toronto. https://hdl.handle.net/1807/107431.

Hossenfelder, Sabine. 2021. "Screams for Explanation: Finetuning and Naturalness in the Foundations of Physics." *Synthese* 198 (S16): 3727–45. https://doi.org/10.1007/s11229-019-02377-5.

Ignatius of Loyola. 1952. *The Spiritual Exercises of St. Ignatius. Based on Studies in the Language of the Autograph*. Translated by Louis J. Puhl. Chicago: Loyola Press.

Ijjas, Anna, Paul J. Steinhardt, and Abraham Loeb. 2014. "Inflationary Schism." *Physics Letters B* 736:142–46. http://dx.doi.org/10.1016/j.physletb.2014.07.012.

———. 2017. "Pop Goes the Universe." *Scientific American* 316 (2): 32–39. https://doi.org/10.1038/scientificamerican0217-32.

Lewis, Geraint F., and Luke A. Barnes. 2016. *A Fortunate Universe: Life in a Finely Tuned Cosmos*. Cambridge: Cambridge University Press.

Lonergan, Bernard. 1992. *Insight: A Study of Human Understanding*. Edited by Frederick E. Crowe and Robert M. Doran. Collected Works of Bernard Lonergan, Volume 3. Toronto: University of Toronto Press.

———. 2017. *Method in Theology.* Edited by Robert M. Doran and John D. Dadosky. Second ed., Revised and Augmented. Collected Works of Bernard Lonergan, Volume 14. Toronto: University of Toronto Press.

Paley, William. 2006. *Natural Theology: Or, Evidences of the Existence and Attributes of the Deity*. Edited by Matthew D. Eddy and David Knight. Oxford: Oxford University Press.

Przywara, Erich. 2014. *Analogia Entis: Metaphysics: Original Structure and Universal Rhythm*. Translated by John R. Betz and David Bentley Hart. Grand Rapids: William B. Eerdmans Publishing Co.

Ratzsch, Del, and Jeffrey Koperski. 2022. "Teleological Arguments for God's Existence." In *The Stanford Encyclopedia of Philosophy,* edited by Edward N. Zalta, Spring 2022 Edition. https://plato.stanford.edu/archives/spr2022/entries/teleological-arguments/.

Simon, Friederich. 2021. "Fine-Tuning." In *The Stanford Encyclopedia of Philosophy,* edited by Edward N. Zalta, Winter 2021 Edition. https://plato.stanford.edu/archives/win2021/entries/fine-tuning/.

Stenger, Victor J. 2011. *The Fallacy of Fine-Tuning: Why the Universe Is Not Designed for Us*. Amherst, New York: Prometheus Books.

Swinburne, Richard. 2010. *Is There a God?* Revised edition. Oxford: Oxford University Press.

Tanner, Kathryn. 1988. *God and Creation in Christian Theology: Tyranny or Empowerment?* Oxford: Blackwell.

Thomas Aquinas. 1888. *Summa Theologiae*. Opera Omnia Iussu Impensaque Leonis XIII P.M., t. IV–XII. Rome: Typographia Polyglotta.

———. 1912–1936. *Summa Theologiae*. Translated by the Fathers of the English Dominican Province. London: Burns, Oates, and Washburne.

———. 1961. *Liber de Veritate Catholicae Fidei Contra Errores Infidelium Seu Summa Contra Gentiles*. Edited by P. Marc, C Pera, and P. Caramello. t. II–III. Taurini-Rome: Marietti.

———. 1963. *Commentary on Aristotle's Physics*. Translated by Richard J. Blackwell, Richard J. Spath, and W. Edmund Thirlkel. New Haven: Yale University Press.

———. 1965. *Quaestiones Disputatae de Potentia*. Edited by P. M. Pession. 10th ed. Quaestiones Disputatae, t. II. Taurini-Rome: Marietti.

Weinberg, Steven. 2007. "Living in the Multiverse." In *Universe Or Multiverse?*, edited by Bernard Carr, 29–42. Cambridge: Cambridge University Press.